\documentclass[superscriptaddress,aps,pra,onecolumn,showpacs,nofootinbib,longbibliography,notitlepage]{revtex4}
\usepackage{etex}
\usepackage{amsfonts}
\usepackage{amssymb}
\usepackage{mathtools}
\usepackage{graphicx}
\usepackage{subfigure}
\usepackage{color}
\usepackage[normalem]{ulem}
\usepackage{natbib}
\usepackage{bbold}
\usepackage{placeins}
\usepackage{braket}
\usepackage{soul,xcolor}
\usepackage{epstopdf}
\usepackage{tabu}
\usepackage{array}
\usepackage{upgreek}
\usepackage{multirow}
\usepackage[utf8]{inputenc}
\usepackage[colorlinks=true,citecolor=blue,urlcolor=blue]{hyperref}
\usepackage{times,txfonts}

\newcommand{\opt}{\operatorname}

\begin{document}
	\title{Quantumness of channels}
	
	\author{Javid Naikoo}
	\email{naikoo.1@iitj.ac.in}
	\affiliation{Indian Institute of Technology Jodhpur, Jodhpur 342011, India}
	
	\author{Subhashish Banerjee}
	\email{subhashish@iitj.ac.in}
	\affiliation{Indian Institute of Technology Jodhpur, Jodhpur 342011, India}
	
	\author{R. Srikanth }
	\email{srik@poornaprajna.org}
	\affiliation{Poornaprajna Institute of Scientific Research, Bangalore - 562164, India}

	\begin{abstract}
	 The reliability of quantum channels for transmitting  information is of profound importance from the perspective of quantum information. This naturally  leads to the question as  how well a quantum state is preserved when subjected to a quantum channel. We  propose a measure of quantumness of channels based on non-commutativity of quantum states that is intuitive and  easy to compute. We apply the proposed measure to some well known noise channels, both Markovian as well as non-Markovian and find that the results are in good agreement with those from a recently introduced $l_1$-norm coherence based measure.
	\end{abstract}
	\maketitle

	\section{Introduction}
	
     Quantifying the degree of quantumness of a channel has both theoretical and practical significance in quantum information science \cite{Horodecki2009quantum}. Quantum channels refer to completely positive and trace preserving maps and can be practically applied to transfer quantum information in a given environment \cite{holevo2012quantum}.    
     To this end, it is important that a channel should preserve the integrity of transmitted quantum states and resist classicalization of the states. Clearly,  it is important to quantify such a degree of quantumness of a quantum channel \cite{Maniscalco}. Considering that classical states are usually identified as those whose correlations can be described in terms of classical probabilities, the quantumness of a channel will indicate how well nonclassical correlations such as entanglement, discord and related quantities \cite{Modi2012theclassical} of a transmitted state are preserved by the channel.  One particularly simple approach to quantifying the quantumness of a channel is to build it on the quantumness measure for a single system, essentially asking how well it preserves the non-commuting property of two states of the system that initially possess this property. This approach has the advantage that it makes no reference to the correlations and requires no complicated optimization procedures \cite{iyengar2013quantifying}.\bigskip
     
Channel noise is usually known for its detrimental role in reducing the degree of coherence in a system, and thus tends to undermine the quantumness of a channel \cite{banerjee2018open}. For example, the deteriorating effect of the environment on a quantum state has been studied in the context of  coherence-breaking channels and coherence sudden death  \cite{Bu2016coherence}. An interesting class of channels known as semi-classical channels $\Lambda_{SC}$  map all the input states $\rho$ to  $\Lambda_{SC}(\rho)$, such that the later are diagonal in the same basis. Such channels are realized by complete decoherence after which only diagonal elements of the density matrix are non-zero \cite{Streltsov2011behavior}. 

However, factors such as squeezing \cite{srikanth2008squeezed,banerjee2008geometric,Inranil} or non-Markovianity \cite{PradeepOSID,Shrikant2018nmd,Li2018concepts} can counterbalance the effect of decoherence for some states, and thus are conducive to the quantumness of the channel.  Refs. \cite{Bddzia2000local,Yeo2008local,Hu2010noise} show that local environments can enhance the average fidelity of quantum teleportation for certain entangled states. Enhancement in quantum discord by local Markovian (i.e., memoryless) noise channels  was reported in \cite{Hu2011necessary,Ciccarello2012creating}.  In \cite{Mani2015cohering}, it was shown that the quantum channels need not be decohering, but could also have cohering power \cite{Kaifeng2017cohering}, which is upper bounded by the corresponding unitary operation. 

Unitary operations may be considered as noiseless channels, and they can give rise to a notion of quantumness based on their entangling power. The entangling capabilities of unitary operations acting on bipartite systems was reported in \cite{Zanardi2000entangling}, with the maximum entanglement being created with  product input states \cite{Shen2018entangling}. 

On the specific question of quantumness of channels, it may be worth noting that in \cite{datta2018coherence, korzekwa2018distinguishing, korzekwa2018coherifying}, coherence of quantum channels was analyzed using Choi-Jamiolwski isomorphism. A coherence based measure of quantumness of channel was proposed in \cite{Shahbeigi2018}, by defining the measure as the average quantum coherence of the state after the quantum channel acts on it, and minimized over all orthonormal basis sets of the state space. This measure was studied in the context of various (non-)Markovian channels \cite{naikoo2019study}.  Further, this measure connects different coherence and entanglement measures, and is also the upper bound for another important coherence measure called \textit{robustness of coherence} for all qubit states \cite{Napoli2016robustness}. 
 
  A necessary and sufficient condition for the creation of quantum correlations via local channels in finite dimensions is that they should not be \textit{commutativity preserving} \cite{Hu2012necessary}. Commutative quantum channels  preserve the commutation relation of any two compatible states, i.e., if $[\rho, \sigma] = 0$, then $[\mathcal{E}(\rho), \mathcal{E}(\sigma)] = 0$. It is  clear that the semiclassical channels, defined above, are commutativity preserving, implying that a departure from semiclassicality is necessary to create quantum  correlations.  In that spirit, here we propose a simple measure for the quantumness of channels, based on commutation properties of the states evolving under a given channel. The degree to which a quantum channel  preserves the non-commutativity of two states can be attributed as the  quantumness of the channel. \bigskip

     This paper is organized as follows: In Sec. \ref{measure} we introduce a measure of quantumness of channels. Section \ref{applications} is devoted to applying this measure to various well-known quantum channels. The experimental relevance of this measure is discussed in Sec. \ref{sec.Exp}. Results and their discussion are presented in Sec. \ref{RandD}. We conclude in Sec. \ref{conclusion}.

      \section{Quantumness of channels}\label{measure}
                  Given two \textit{arbitrary} states $\rho$ and $\sigma$, one can quantify their mutual incompatibility by the Hilbert-Schmidt (HS) norm of their commutator   $\mathcal{M}(\rho, \sigma) =  2 ||  \opt{C}||^{2}_{HS} \equiv 2 \operatorname{Tr}[\opt{C}^\dagger \opt{C}]$. The measure is defined in terms of the HS  norm of their commutator $\opt{C} = \rho \sigma - \sigma \rho$.  This measure was motivated in \cite{iyengar2013quantifying}  with the aim of identifying nonclassicality with the incompatibility of states.   Consider two qubit states $\rho_a = \frac{1}{2}(\mathbf{1} + \vec{a} \cdot \vec{\sigma})$ and $\rho_b = \frac{1}{2}(\mathbf{1} + \vec{b} \cdot \vec{\sigma})$, with $\vec{a}, \vec{b} \in \mathbb{R}^3$ and $\vec{\sigma} = (\sigma_x, \sigma_y, \sigma_z)$ represents the three Pauli spin matrices. We have  $ \rho_a \rho_b - \rho_b \rho_a = i \frac{1}{2} (\vec{a} \times \vec{b}) \cdot \vec{\sigma}$, and 
                             \begin{equation}
                                    \mathcal{M}(\rho_a, \rho_b)  = 2 \opt{Tr}\{( \rho_a \rho_b - \rho_b \rho_a)^\dagger ( \rho_a \rho_b - \rho_b \rho_a)\} = | \vec{a} \times \vec{b}|^2.
                             \end{equation}
                              This quantity attains its maximum value of one for orthogonal $\vec{a}$ and $\vec{b}$ and  vanishes when  $\vec{a}$ and $\vec{b}$  are  parallel, $0 \le \mathcal{M}(\rho, \sigma) \le 1$.
              \bigskip

           Here we try to exploit this approach to probe  the quantumness of a channel. Consider a channel described by a linear, completely positive and trace preserving map $\Phi: L(\mathcal{H}_A) \rightarrow  L(\mathcal{H}_B)$ \cite{watrous2018theory,wilde2013quantum}. The action of this map on an input state $\rho$ leads to an output state $\rho^\prime$ and can be summarized as 
          \begin{equation}\label{eq:map}
          \rho^\prime = \Phi[\rho].
          \end{equation}
          In the context of quantum channels, we start with two states $\rho_a$ and $\rho_b$ which are  maximally noncommuting in the sense that $\mathcal{M}(\rho_a, \rho_b) = 1$.  By  subjecting these  states to a quantum channel, the quantumness of the channel can be attributed to  the extent to which $\rho_a^\prime$ and $\rho_b^\prime$ (the outputs) are incompatible  
          \begin{equation}\label{eq:Mrhoarhobprime}
          \mathcal{M}(\rho_a^\prime, \rho_b^\prime) =  2 \operatorname{Tr}[\opt{C}^\dagger \opt{C}], 
          \end{equation}
           with $\opt{C} = \rho_a^\prime \rho_b^\prime - \rho_b^\prime \rho_a^\prime$.  This quantity when maximized over all input states serves as a measure for the quantumness of the channel
           \begin{equation}
           \mu = \max_{\rho_a,~ \rho_b}  \mathcal{M}(\rho_a^\prime = \Phi[\rho_a], \rho_b^\prime = \Phi[\rho_b]).
           \end{equation}
             As an example, consider the states $\ket{a}= \cos(x/2) \ket{0} + e^{-i \phi} \sin(x/2) \ket{1}$ and $\ket{b} = \cos(y/2) \ket{0} + e^{-i \xi } \sin(y/2) \ket{1}$, with the respective density matrix representations
          \begin{align}\label{rhoarhob}
          \rho_a &=\frac{1}{2} (\mathbf{1} + \vec{a}\cdot \vec{\sigma}) = \begin{pmatrix}
                              \cos^2(x/2) & e^{i \phi} \frac{\sin(x)}{2}\\
                              e^{-i \phi}\frac{\sin(x)}{2} & \sin^2(x/2)
                          \end{pmatrix}, \nonumber \\ 
          \rho_b &=\frac{1}{2} (\mathbf{1} + \vec{b}\cdot \vec{\sigma}) =   \begin{pmatrix}
                                                                                             \cos^2(y/2) & e^{i \phi} \frac{\sin(y)}{2}\\
                                                                                               e^{-i \xi}\frac{\sin(y)}{2} & \sin^2(y/2)
                                                                                           \end{pmatrix}.
          \end{align}
          The states $\rho_a$ and $\rho_b$ are maximally non-commuting for $y = x + \pi/2$ and $\xi = \phi$, as can be seen by calculating  the commutator
          \begin{equation}
          \opt{C} = \rho_a \rho_b - \rho_b \rho_a = \begin{pmatrix}
                                                                                0  &  \frac{e^{i \phi}}{2}\\
                                                                               - \frac{e^{-i \phi}}{2} & 0            
                                                                        \end{pmatrix}.
          \end{equation} 
          Therefore $\mathcal{M}(\rho_a, \rho_b) =  2  \opt{Tr}[\opt{C}^\dagger \opt{C}] = 1$. Thus the states are maximally noncommuting and in this sense share maximum nonclassicality.   In this example, no optimization is required since the quantity $\mathcal{M}(\rho_a, \rho_b)$ is independent of input state parameters. However, as discussed ahead, subjecting these states to quantum channels can make $\mathcal{M}(\rho_a, \rho_b)$ dependent on input state parameters. In such cases, we need to maximize over all such parameters to compute the degree of incompatibility of the output states.
      \section{Application to quantum channels} \label{applications}
        We will now apply the above definition to some well known quantum channels. We consider the dephasing channels like random telegraph noise (RTN) \cite{Daffer}, non-Markovian dephasing (NMD) \cite{Shrikant2018nmd}, phase damping (PD) \cite{banerjee2007dynamics} and generalized depolarizing channel (GDC) \cite{nielsen2002quantum}. The generalized amplitude damping channel (GAD) \cite{srikanth2008squeezed,omkar2013dissipative}, which represents a dissipative channel is also studied.  The Kraus operators for these channels are given in Table (\ref{tabquantumness}).

        \bigskip

      Random Telegraph Noise (RTN):
        The dynamical map is represented by the Kraus operators $K_0(t) = k_+ ~ \mathbf{I}$ and  $K_1(t) = k_{-}  ~ \sigma_z$, where $k_{\pm} = \sqrt{ \frac{1 \pm  \Lambda (t)}{2}}$, such that the action on a general qubit state 
        \begin{equation}
        \rho = \begin{pmatrix}
                        1-p  &  x\\
                     x^*   & p
                    \end{pmatrix},
        \end{equation}
        is given by
        \begin{equation}
        \rho^\prime = \Phi^{RTN} \begin{pmatrix}
                                                        1-p  &  x\\
                                                       x^*   & p
                                             \end{pmatrix}
                            =                          \begin{pmatrix}
                                                                1-p  & x  \Lambda (t) \\
                                                                 x^* \Lambda (t)    & p
                                                         \end{pmatrix}.
                                                     \end{equation}
           Let us use the maximally nonclassical pair of states given in Eq. (\ref{rhoarhob}), subject to the constraints enunciated below it. The  states $\rho_a$ and $\rho_b$, defined in Eq. (\ref{rhoarhob}) are subjected to RTN evolution                                 
         \begin{align}\label{eq:rhobRTN}
        \rho^\prime_a &=\left(
        \begin{array}{cc}
         \cos ^2\left(\frac{x}{2}\right) & \frac{1}{2} e^{i \phi } \sin (x) \Lambda (t) \\
        \frac{1}{2} e^{-i \phi } \sin (x) \Lambda (t) & \sin ^2\left(\frac{x}{2}\right) \\
        \end{array}
        \right),\nonumber \\
        \rho^\prime_b &=  \left(
        \begin{array}{cc}
        \cos ^2\left(\frac{1}{4} (2 x+\pi )\right) & \frac{1}{2} e^{i \phi } \cos (x) \Lambda (t) \\
        \frac{1}{2} e^{-i \phi } \cos (x) \Lambda (t) & \sin ^2\left(\frac{1}{4} (2 x+\pi )\right) \\
        \end{array}
        \right).
        \end{align}   
        The pertinent commutator in this case becomes 
        \begin{equation}\label{eq:CRTN}
        \opt{C} =\left(
        \begin{array}{cc}
        0 & \frac{1}{2} e^{i \phi } \Lambda (t) \\
        -\frac{1}{2} e^{-i \phi } \Lambda (t) & 0 \\
        \end{array}
        \right).
        \end{equation}
        Therefore, the quantumness measure for the RTN channel turns out to be    $\mu = \max_{\rho_a,~ \rho_b} \mathcal{M}(\rho_a^\prime, \rho_b^\prime) =  2  \opt{Tr}[\opt{C}^\dagger \opt{C}] = [\Lambda(t)]^2$.  Similarly, for non-Markovian dephasing (NMD) and phase damping (PD) channels, characterized by the Kraus operators in Table (\ref{tabquantumness}), the quantumness measure turns out to be
        \begin{align}
          NMD: ~ \mu &= \max_{\rho_a,~ \rho_b} \mathcal{M}(\rho_a^\prime, \rho_b^\prime) = [\Omega(p)]^2, \label{eq:NMD} \\
          PD:  ~ \mu &= \max_{\rho_a,~ \rho_b} \mathcal{M}(\rho_a^\prime, \rho_b^\prime) = 1-\gamma.\label{eq:PD}
        \end{align}

   It may be pointed out that the channels RTN, NMD and PD are all types of dephasing channels, and hence they are not expected to show dissimilar behavior. However,  they are all included, partly for two reasons: (a) comparison with previous literature reporting on the quantumness of these channels; (b) they illustrate different specific principles of decoherence, and thus indicate distinct underlying physics. Specifically, RTN is a non-Markovian (P-indivisible) channel having infinitely many singularities, and NMD is a non-Markovian (also, P-indivisible) channel with a single singularity, whereas PD can be Markovian.

    	\bigskip

     Next, we consider an example of non-dephasing class of quantum channel, namely the amplitude damping (AD) channel. This channel  models the processes like spontaneous emission and is characterized by Kraus operators $A_0$, $A_1$ given in Table (\ref{tabquantumness}). Here $\gamma \in [0,1]$. The action of AD channel on a qubit state can be described as
    		\begin{equation}\label{eq:ADactioin}
    		\begin{pmatrix}
    	1-p  ~& ~q\\\\
    	q^* ~ &~ p
    		\end{pmatrix}                                                \rightarrow  	\begin{pmatrix}
    	                                                                                                         	1-p(1-\gamma) ~&  ~q \sqrt{1-\gamma}\\\\
    		                                                                                                        q^* \sqrt{1-\gamma}                    ~& ~                  p(1-\gamma)
    		                                                                                                            \end{pmatrix}  
    		\end{equation}    		
    		 Following the recipe given above, the quantumness in this case turns out to be
\begin{equation}\label{eq:AD}
\mu = \max_{\rho_a,~ \rho_b} \mathcal{M}(\rho_a^\prime, \rho_b^\prime)  = 1 - \gamma.
\end{equation}    
It follows that as $\gamma \rightarrow 1$, the quantumness parameters becomes zero, and the state looses coherence in the given basis, see Eq. (\ref*{eq:ADactioin}).   

\bigskip

 We next consider the example of the Unruh channel. The Unruh effect is the emergence of finite temperature recorded by an observer undergoing acceleration $a$ in a vacuum bath. Yet, paradoxically, it is a rank 2 (rather than full rank) noise effect and in that sense is similar to the AD channel, which represents the effect of a zero temperature thermal bath. The parameter $r$ appearing in Kraus operators is related to acceleration by $\cos(r) = (1 + e^{-2 \pi \omega c/a})^{-1/2}$, where $\omega$ is the Dirac particle frequency and $c$ is speed of light in vacuum. The quantumness parameter for this channel is given by 
\begin{equation}\label{eq:Unruh}
\mu = \cos^2(r) = (1 + e^{-2 \pi \omega c/a})^{-1},
\end{equation}
and  approaches one in the limit $a \ll 2\pi c~ \omega$.

    \bigskip

   Our final example is the generalized depolarizing channel (GDC), represented by the following Kraus operators  $M_i = \sqrt{p_i} \sigma_i$ with $i = 0,1,2,3$, where $\sigma_i$ are the Pauli matrices.  The states $\rho_a$ and $\rho_b$ given by Eq. (\ref*{rhoarhob}) evolve under the action of this channel such that the new Bloch vectors are given by
\begin{align}
\vec{a} &= \begin{pmatrix}
      (p_0 + p_1 - p_2 - p_3) \sin x \cos \phi \\
      (-p_0 + p_1  - p_2 + p_3) \sin x \sin \phi \\
      (p_0 - p_1 - p_2 + p_3) \cos x
\end{pmatrix}, \nonumber \\
\vec{b} &= \begin{pmatrix}
                       (p_0 + p_1 - p_2 - p_3) \cos x \cos \phi \\
                       (-p_0 + p_1 - p_2 + p_3) \cos x \sin \phi \\
                        (-p_0+p_1+p_2-p_3) \sin x
\end{pmatrix}.
\end{align}
Therefore,
\begin{align} \label{eq:MGDC}
\mathcal{M}(\rho_a^\prime , \rho_b^\prime) &= |\vec{a} \times \vec{b}|=
                            (p_0 - p_1 - p_2 + p_3)^2 [2 (p_0 - p_3) (p_1 - p_2) \cos(2\phi) + (p_0 - p_3)^2 + (p_1 - p_2)^2].
\end{align}
This is maximum for $\phi = 0$, i.e., $\mu =  \mathcal{M}(\rho_a^\prime , \rho_b^\prime)|_{\phi = 0} = (p_0 + p_1 -p_2 -p_3)^2 (p_0 - p_1 - p_2 + p_3)^2$.   The various results discussed above are summarized in Table (\ref{tabquantumness}), and compared with the prediction of a coherence-based measure of the quantumness of channels.

\section{Experimental relevance of the measure}\label{sec.Exp}
 It is important to note that the quantity $\mathcal{M}(\rho_a^\prime, \rho_b^\prime)$ can be given an experimental interpretation using an interferometric setup \cite{ferro2015measuring}. This useful technique can be easily incorporated to our purpose of quantifying quantumness of channels. One can write 
\begin{equation}\label{eq:Mtrace}
\mu = \max_{\rho_a,~ \rho_b} \mathcal{M}(\rho_a^\prime, \rho_b^\prime)  = 4  \max_{\rho_a,~ \rho_b} \opt{Tr}[(\rho_a^\prime)^2 (\rho_b^\prime)^2 - (\rho_a^\prime \rho_b^\prime)^2 ].
\end{equation} 
 The two quantities  $\opt{Tr}[(\rho_a^\prime)^2 (\rho_b^\prime)^2]$ and $\opt{Tr}[(\rho_a^\prime \rho_b^\prime)^2]$ can be obtained from two separate measurements. The input state $\rho = |0\rangle \langle 0 |  \otimes \rho_a^\prime \otimes \rho_a^\prime \otimes \rho_b^\prime \otimes \rho_b^\prime$, where $\ket{0}$ is the control qubit, is subjected to the controlled unitary gate $U$.  This modifies the interference of the controlled qubit by the factor $\opt{Tr}[\rho U] = v e^{i \alpha}$, with $v$ and $\alpha$ being the visibility and phase shift of the interference fringes, respectively \cite{Sjoqvist2000geometric,Filip2002overlap,Carteret2005noiseless,Ekert2002direct}.  Two such schemes (corresponding to  $\opt{Tr}[(\rho_a^\prime)^2 (\rho_b^\prime)^2]$ and $\opt{Tr}[(\rho_a^\prime \rho_b^\prime)^2]$ ) lead to the  quantumness $\mathcal{M}(\rho_a^\prime, \rho_b^\prime) = 4 (v_1 - v_2)$, where $v_1$ and $v_2$ correspond to the respective visibilities obtained by the action of relevant unitary gates.  We motivate the present discussion by illustrating this notion on some of the channels discussed above.

 (a)  For RTN, the two visibilities (with $x=\phi=0$) correspond to  
 \begin{equation}
 \opt{Tr}[(\rho_a^\prime \rho_b^\prime)^2] = \frac{1}{4}, \quad 
  \opt{Tr}[(\rho_a^\prime)^2 (\rho_b^\prime)^2 ] = \frac{1}{4}(1 + [\Lambda(t)]^2).
 \end{equation}
   Making use of these in Eq. (\ref{eq:Mtrace}), we obtain $\mathcal{M}(\rho_a^\prime, \rho_b^\prime) = [\Lambda(t)]^2$, consistent with the definition in Eq. (\ref{eq:Mrhoarhobprime}), see below Eq. (\ref{eq:CRTN}).

  (b)  For GDC, the two visibilities (with $x=\phi=0$) turn out to be 
 \begin{align}
 \opt{Tr}[(\rho_a \rho_b^\prime)^2] &= 1/4 - 2 (-1 + p_1 + p_2) (p_1 + p_2) (-1 + p_2 + p_3) (p_2 + p_3),\nonumber\\
 \opt{Tr}[(\rho_a^\prime)^2 (\rho_b^\prime)^2 ] &= 1/2 (1 + 2 p_1^2 + 2 (-1 + p_2) p_2+ p_1 (-2 + 4 p_2)) (1 + 2 p_2^2   \nonumber \\&+ 
 2 (-1 + p_3) p_3 + p_2 (-2 + 4 p_3)).
 \end{align}   
 These lead to the expression    $\mu = \mathcal{M}(\rho_a^\prime, \rho_b^\prime)|_{\phi =0} =  (p_0 + p_1 -p_2 -p_3)^2 (p_0 - p_1 - p_2 + p_3)^2$ in accord with the definition in Eq. (\ref{eq:Mrhoarhobprime}), see Eq. (\ref{eq:MGDC}). 
 
 What makes this approach particularly attractive is that here the quantumness of the channel can be experimentally determined.

        		\begin{table}[h!]
        		\centering
        		\caption{\label{tabquantumness} Various quantum channels, introduced at the beginning of Sec. \ref{applications},  with their Kraus operators and the quantumness using commutation based measure $\mu = \max_{\rho_a, ~\rho_b} \mathcal{M}(\rho_a^\prime, \rho_b^\prime)$. For the sake of comparision,  the corresponding results based on the  coherence based measure $Q_{C_{l_1}}$ \cite{naikoo2019study} are also provided. Here, $\tilde{\xi} = \frac{5}{2}(\alpha - 1)^2 (1-\xi)^2$ and $\tau = \frac{-2}{\gamma (2n+1)} \ln \big[ \frac{5}{6+4n + n^2}\big]$ \cite{naikoo2019study}. }
        		\label{summary}
        \begin{tabular}{ |p{1cm}|p{6cm}|p{2.4cm}|p{2cm}|}
        	\hline
        	Channel                                         & Kraus operators &  $\mu$ & $Q_{C_{l_1}}$  \\
        	\hline
        	RTN & \small $K_0 = k_+ \begin{pmatrix} 1 & 0\\ 0 & 1 \end{pmatrix}$, $K_1 = k_- \begin{pmatrix} 1 & 0\\ 0 & -1 \end{pmatrix}$. \normalsize & $[\Lambda(t)]^2$ & $[\Lambda(t)]^2$ \\
        	\hline
             NMD &  \small $N_0 = n_+ \begin{pmatrix} 1 & 0\\ 0 & 1 \end{pmatrix}$,$ N_1 = n_- \begin{pmatrix} 1 & 0\\ 0 & -1 \end{pmatrix}$. \normalsize    & $[\Omega(p)]^2$ & $[\Omega(p)]^2$ \\
        	\hline
        	PD & \small $ P_0 = \begin{pmatrix}
        	1 &  0 \\
        	0  &  \sqrt{1-\gamma}
        	\end{pmatrix}$, $P_1 = \begin{pmatrix}
        	1 & 0\\
        	0 & \sqrt{\gamma}
        	\end{pmatrix}$. \normalsize & $1-\gamma$ & $1-\gamma$ \\
        	\hline
        	Unruh    & \small $U_0 = \begin{pmatrix}
        		\cos(r)  &   0\\
        		0          &  1        
        	\end{pmatrix}$, $U_1 = \begin{pmatrix}
        		0  &   0\\
        		\sin(r)   &  0        
        	\end{pmatrix}$. \normalsize & $\cos^2(r)$  &   $\cos^2(r)$ \\
        	\hline
        	AD &  $A_0 = \small \begin{pmatrix} 1 & 0\\ 0 & \sqrt{1- \gamma} \end{pmatrix}$, $A_1 = \begin{pmatrix} 0 & \sqrt{\gamma}\\ 0 &0   \end{pmatrix}$ \normalsize  & $1- \gamma$ & $1- \gamma$ ($\gamma > \frac{1}{6}$) \vspace{0.4cm} \newline $\frac{1}{6}(6 \gamma^2  - 3 \gamma +2)$ ($\gamma \le \frac{1}{6}$) \\
        	\hline
             GAD & $G_0 = \small \begin{pmatrix} \sqrt{\alpha} & 0\\ 0 & \sqrt{\alpha \xi} \end{pmatrix}$, $G_1 = \begin{pmatrix} 0 & \sqrt{\alpha P}\\ 0 & 0 \end{pmatrix}$, \small $G_3 = \begin{pmatrix} \sqrt{\beta \xi} & 0\\ 0 & \sqrt{\beta}  \end{pmatrix}$\normalsize, \small $G_4 = \begin{pmatrix} 0 & 0\\ \sqrt{\beta P} & 0  \end{pmatrix}$.\normalsize  \normalsize& $\xi  \left(\xi-\sqrt{2} (\xi-1)\right)^2$  ($\xi>1$) \newline $\xi  (-1 + 2 \xi)^2$  ($\xi <1$)  &  $\xi$ ($t>\tau$) \vspace{0.5cm } \newline $\frac{1}{2} \xi + \tilde{\xi}$ ($t \le \tau$) \\
        	\hline
        \end{tabular}
    \end{table}

\section{Results and discussion}\label{RandD}  
The quantumness of two arbitrary states $\rho$ and $\sigma$ can be identified with their incompatibility and quantified by $\mu = \max_{\rho, ~\sigma} \mathcal{M}(\rho, \sigma)$ as defined in Sec. \ref{measure}.  For a mixed initial diagonal state $\rho_0 = \sum_{i} \lambda_i | i \rangle \langle i |$, which evolves to $\rho_t$ under some dynamics,  the following inequality holds \cite{bhattacharya2018evolution}
\begin{equation}\label{eq:ineq}
\frac{\mathcal{M}(\rho_0, \rho_t) }{4}   \le F(\rho_0, \rho_t)   \le \frac{C_{l_1}(\rho_0, \rho_t)}{2}.
\end{equation}
Here, $F(\rho_0, \rho_t)$ is the quantum Fisher information and $C_{l_1}(\rho_0, \rho_t)$ is the $l_1$-norm coherence of   $\rho_t$, defined as $\sum_{i\ne j} \lvert \rho_t^{ij}\rvert$; both  well known measures of  quantumness. The commutator based measure provides a lower bound and a reliable witness of quantumness.  \bigskip

In this work, we extend the approach of quantifying the quantumness of  states,  in terms of their incompatibility, to explore the  quantumness of channels. This method involves   starting with two states which are maximally non-commuting and subjecting them to a quantum channel. The incompatibility of the resulting output states can be attributed to the  degree of quantumness of the channel.  We have  computed the quantumness of various well known channels and compared them with the analogous estimation of quantumness from a coherence based measure \cite{Shahbeigi2018}. These are listed  in Table (\ref{tabquantumness}).  Specifically, we investigated the quantumness of the dephasing channels such as RTN, non-Markovian dephasing (NMD), phase damping (PD), and Unruh channels. These channels model the phenomenon of decoherence without dissipation. The dissipative channels considered here are amplitude damping (AD) and generalized amplitude damping  (GAD) channels. The quantumness is given in terms of the channel parameters. In particular, for the channels with memory, the quantumness turns out to be a function of the memory kernel, which in turn, decides the Markovian and non-Markovian nature of the dynamics. As reported in \cite{naikoo2019study}, the non-Markovian dynamics helps to sustain the quantumness over longer time as compared to Markovian case. It is interesting to note that quantumness from the proposed measure is in good agreement with that with the coherence based measure \cite{naikoo2019study}. This is consistent with our intuition as coherence is related to the off-diagonal elements of the density matrix as would be the cause for noncommutativity between the states.  It should be noted that in the case of GDC,  the coherence based measure  leads to  quantumness   $(p_0 -p_1)^2 + (p_2 - p_3)^2$, different from that obtained by the commutation based measure adapted here. This is consistent with Eq. (\ref{eq:ineq}).

The attractive feature here is that the measure proposed   can be calculated easily and is also amenable to experimental determination. Further, from the cases of the RTN and NMD channels, it  is evident that quantumness reflects the non-Markovian nature of the channel under consideration.\bigskip

\section{Conclusion}\label{conclusion}
The quantum channels provide a way to describe the processes where pure states go over to mixed ones.  Therefore, it is natural to ask how well a quantum channel preserves the quantumness of the states which are subjected to it. Recently, a measure based on the $l_1$-norm coherence was introduced to quantify the quantumness of channels. In this work, we have addressed the  problem by using an intuitive approach based on the incompatibility of the states.  The quantumness of a system is identified with the  mutual non-commutation of all its accessible states.  We illustrated the approach developed here by considering various examples of quantum channels, both Markovian as well as non-Markovian, and found that our results are in good agreement with the coherence based measure. An added attraction of this method is that it  is easy to compute and  can be probed experimentally.

%
\end{document}